\documentclass[a4paper,fleqn,usenatbib]{mnras}

\usepackage{mathptmx}
\usepackage{txfonts}

\usepackage[T1]{fontenc}
\usepackage{ae,aecompl}

\usepackage{graphicx}	
\usepackage{amssymb}	
\usepackage{hyperref}
\usepackage{longtable}

\newcommand{\jgrl}{Geophys. Res. Lett.}
\newcommand{\jgrp}{J. Geophys. Res. Planets}

\newcommand{\jgrs}{J. Geophys. Res. Space Phys.}
\newcommand{\jpca}{J. Phys. Chem. A}
\newcommand{\epscl}{Earth Planet. Sci. Lett.}

\title[$^{12,13}$C+CO$_2$ scattering cross sections]{Elastic and inelastic cross sections for $^{12}$C+CO$_2$ and $^{13}$C+CO$_2$ scattering at superthermal energies}

\author[M. Gacesa]{
Marko Gacesa,$^{1,2,3}$\thanks{email: marko.gacesa@ku.ac.ae}
\\
$^{1}$Department of Physics, Khalifa University, Abu Dhabi, United Arab Emirates \\
$^{2}$Center for Catalysis and Separation (CeCaS), Khalifa University, Abu Dhabi, United Arab Emirates \\
$^{3}$Bay Area Environmental Research Institute, Moffett Field, CA, USA 
}

\pubyear{2023}

\begin{document}
\label{firstpage}
\pagerange{\pageref{firstpage}--\pageref{lastpage}}
\maketitle

\begin{abstract}
We report elastic and inelastic cross sections for fast superthermal $^{12}$C($^3P$) and $^{13}$C($^3P$) atoms scattering on $^{12}$CO$_2$. The cross sections were computed using quantum-mechanical rotationally close-coupling formalism with the electronic interaction described by a newly constructed potential energy surface correlating to the lowest energy asymptote of the complex. State-to-state cross sections, differential cross sections, and derived transport properties of interest for energy relaxation are also reported. 
The computed elastic cross sections are strongly anisotropic, show significant energy dependence, and differ by up to 2\% between the two isotopes of carbon.
\end{abstract}




\section{Introduction}

Carbon is one of the most abundant elements in the Universe and an active participant in chemistry of dense interstellar clouds in star-forming regions, protostellar disks, and (exo)planetary atmospheres \citep{2013RvMP...85.1021T,sandford2020prebiotic}. There is evidence that chemical reactions of hot C($^3$P) atom with interstellar ices, such as CO$_2$ ice found in protoplanetary disks, produce complex organic molecule \citep{2011ppcd.book...55B,2017A&A...608A..12S} that could be brought to surfaces of newly-formed planetary bodies where they might play a key role in the origin of life. 

In atmospheres of terrestrial planets and exoplanets, carbon dioxide acts as a significant coolant molecule \citep{sharma1990role} and participates in photochemical reactions driven by stellar radiation and stellar wind.
In CO$_2$-dominated atmospheres of Mars and Venus, photochemistry of CO and CO$_2$ molecules driven by UV solar photons produces hot carbon atoms in both ground and excited electronic states, some of which can have translational energies as high as several eVs, making them strongly superthermal with respect to the background atmospheric gases \citep{2001JGR...10628785F,2021Icar..36014371L}. At Mars, the photochemical processes appear to be a major driver of non-thermal atmospheric escape of atomic species to space, estimated to be responsible for the loss of hundreds of millibars of CO$_2$ in the present epoch \citep{2018Icar..315..146J,jakosky2023mars}. 
Photochemical escape of atomic oxygen, where the production of energetic O atoms largely occurs through dissociative recombination of O$_{2}^{+}$ with electrons, has long been regarded as the dominant such process \citep{2009Icar..204..527F,2017JGRA..122.3815L,2018Icar..300..411F} and studied extensively in the context of climate evolution on Mars and recently confirmed by NASA's Mars Atmosphere and Volatile EvolutioN (MAVEN) mission \citep{2015GeoRL..42.9009D,2017JGRA..122.3815L,2017JGRE..122.2401L}. 
Similarly, photochemical escape of atomic carbon is regarded as a possible explanation for a fraction of the observed deficit of carbon-bearing minerals at Mars \citep{2001JGR...10628785F,tian2009thermal,2014P&SS...98...93G,2019A&A...621A..23C,2021Icar..36014371L,2023PSJ.....4...53Y}, especially since the isotopes of Xe do not offer too much insight into the ancient mechanisms of carbon escape \citep{zahnle1993xenological,cassata2022xenon}. Measurements of carbon isotope ratios, such as $^{13}$C/$^{12}$C, carry additional information about the evolution of planetary atmospheres \citep{2007Icar..192..396K,2015NatCo...610003H,2023PSJ.....4...97A} and other carbon-rich cosmic regions \citep{2019ApJ...887..143S}.

In these environments, the collision cross sections, excitation rates, and derived transport properties of superthermal carbon atoms and their isotopes against a thermal CO$_2$ background play an important role. These quantities are not available in the literature for the kinetic energies of several eV. The closest related studies focused on chemical kinetics and reaction rates of the C($^1$D)+CO$_2$ reaction at temperatures between 50 K and 300 K \citep{2018JPCA..122.4002N} and found them to be in good agreement with previously reported measurements \citep{TF9716703166}. 
Consequently, for purposes of modeling astrophysical systems and planetary atmospheres, $^{12,13}$C($^3$P)+CO$_2$ cross sections are commonly estimated by mass-scaling elastic and inelastic cross sections for better known pairs, such as O($^3P$)+CO$_2$ \citep{2020MNRAS.491.5650G}, O+H$_2$ \citep{2014JChPh.141p4324G}, O+N$_2$ \citep{1998JGR...10323393B}, or O+O \citep{2000JGR...10524899K}. A possible mass-scaling approach based on a number of atom-molecule cross sections known at eV energy range and capable of estimating differential cross sections has been proposed by \citet{2014ApJ...790...98L} and adopted in some studies. Nevertheless, such approaches generally cannot reliably estimate the cross sections and derived observables, nor can they reproduce reliable inelastic or state-to-state cross sections for internal excitations of rotational and vibrational modes of carbon dioxide molecule, resulting in significant uncertainties in atmospheric escape models \citep{2000JGR...10524899K,2011GeoRL..3802203B,2014Icar..228..375F,2018Icar..300..411F,lee2020effects}.


In this article, we present a fully quantum-mechanical study of C($^3P$) + CO$_2$ scattering on a newly constructed potential energy surface correlating to the electronic ground state of the complex for two isotopes of carbon, $^{12}$C and $^{13}$C. The cross sections and derived quantities were computed for the collision energies between 0.03 eV and 5 eV. The anticipated areas of application of the results are in modeling of astrophysical environments where non-thermal C atoms are produced, such as planetary and cometary atmospheres, dense photo-dissociating regions, and protoplanetary disks \citep{2019BAAS...51c.240K}.

\section{Methods}

\subsection{Electronic interactions of \texorpdfstring{C($^3P$)-CO$_2$}{O(3P)-CO2} complex}

\begin{figure}
\centering
\noindent\includegraphics[width=0.6\columnwidth]{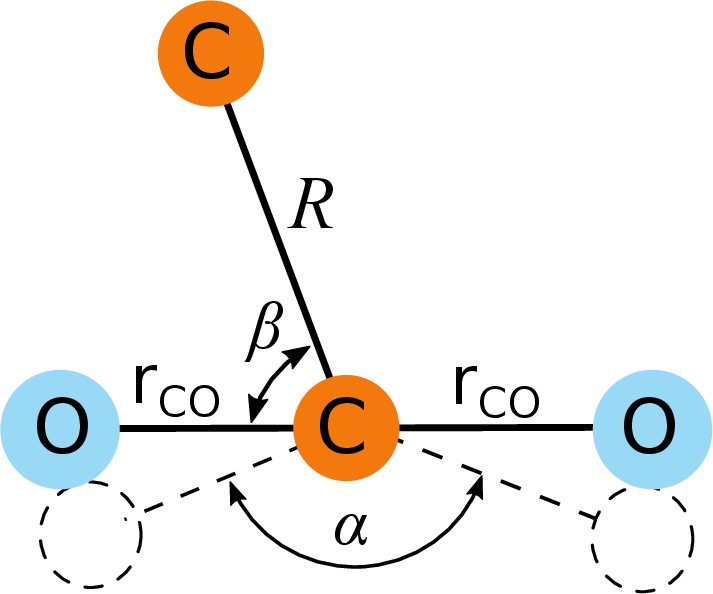}
\caption{Coordinate system used to parameterize the potential energy surfaces of the C+CO$_2$ complex. Planar symmetry is assumed. The bending angle $\alpha=180^\circ$ and $r_\mathrm{CO}=1.167$ $\AA{}$ at equilibrium distance for a non-vibrating CO$_2(00^{0}0)$ molecule.}
\label{fig:geometry}
\end{figure}

In order to compute quantum-mechanical cross sections for C($^3P$) + CO$_2$ scattering, we follow the approach used in our earlier study of O($^3P$) + CO$_2$, a complex with a similar molecular symmetry \citep{2020MNRAS.491.5650G}. 
We first construct the lowest-energy electronic potential energy surface (PESs) correlating to the lowest asymptote of the C$(^3P)-\mathrm{OCO}$ complex. 
It is worth mentioning that we are only interested in the electronic interaction of the complex and not in the ground state configuration of C$_2$O$_2$ molecule: ethylenedione, C$_2$O$_2$, is a rather curious molecule that has eluded attempts at synthesis for more than a century \citep{staudinger1913oxalylchlorid,2019PCCP...2112986K}. Regardless of being an oxocarbon with a closed-shell Kekul\'e structure (O=C=C=O), its ground state is a triplet, $^3\Sigma_{g}^{-}$ \citep{Schroeder1998,bao2012molecular,doi:10.1063/1.4976969}, which makes it a transient long-lived diradical whose electronic configuration is similar to O$_2$ molecule.

We further assume that the target CO$_{2}$ molecule remains in its ground vibrational state, CO$_{2}(0 0^{0} 0)$=CO$_{2}(v_s v^{l}_b v_a)$, where $v_s=0$, $v^{l}_b=0$ and $v_a=0$ are symmetric, bending, and asymmetric vibrational quantum numbers, respectively. Thus, in our model, CO$_{2}(0 0^{0} 0)$ molecule is linear and symmetric, with both C-O internuclear distances fixed at its equilibrium value, $r_{\mathrm{CO}}$ = 1.162 \AA{} \citep{NIST_comput_chemistry_db}.
The rigid rotor assumption preserves the rotational symmetry of the scattering complex with respect to the O-C-O internuclear axis and allows us to use a two-dimensional (2D) potential energy surface to model the electrostatic interactions.
An intuitive parametrization of the PES for purposes of scattering is given by the internuclear distance $R$, defined as the distance between the impacting C atom and the C atom of the CO$_2$ molecule, and the angle $\beta$ between the CO$_2$ internuclear axis and the vector $R$, defined such that $0^\circ \leq \beta \leq 90^\circ$ (Fig. \ref{fig:geometry}). Finally, the bending angle $\alpha$ is set to $\alpha=180^\circ$. 
Thus, the symmetry group of the system is at least $C_S$ and depends on the angle of approach $\beta$. 
The special cases when the symmetry of the system is higher are the T-approach ($\beta=0^\circ$, symmetry group C$_{2v}$), and the collinear geometry ($\beta = 90^\circ$, symmetry group C$_{\infty v}$). 

We computed the electronic potential energies of the complex at the single-reference locally-correlated coupled-cluster singles and doubles with perturbative triples (LCCSD(T)) level of theory on restricted open-shell local orbitals (ROHF) with the frozen core approximation \citep{2011JChPh.135j4111R,2013JChPh.139i4105R,doi:10.1021/acs.jctc.8b00442} on Dunning's aug-cc-pV5Z correlation-consistent basis sets \citep{1989JChPh..90.1007D,1992JChPh..96.6796K}. We settled for this particular choice of the method to obtain a satisfactory computational accuracy and numerical stability while keeping the computational costs manageable. To benchmark the accuracy of the method, we repeated the calculation using a partially spin-restricted explicitly-correlated theory with MP2-F12 approximation (RCCSD(T)-F12) \citep{knizia2009simplified} with the cc-pCVTZ-F12 basis set \citep{2010JChPh.132e4108H}. All calculations were carried out using the MRCC suite of \textit{ab initio} and density functional quantum chemistry programs \citep{kallay2016mrcc,2020JChPh.152g4107K}, configured to use the recent implementation of the local correlation algorithms \citep{nagy2016integral,doi:10.1021/acs.jctc.8b00442}.

\begin{figure}
\centering
\noindent\includegraphics[width=0.95\columnwidth]{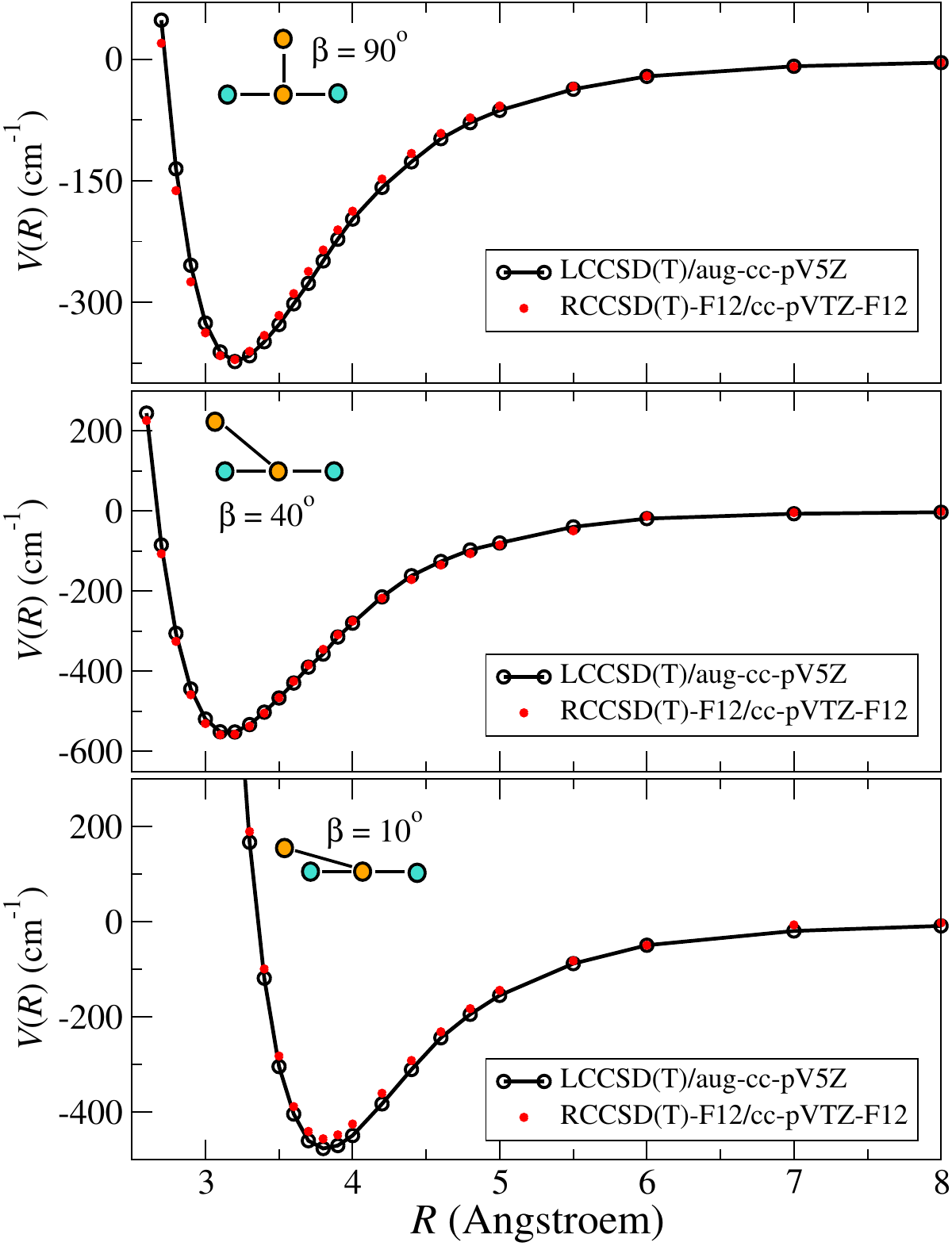}
\caption{\textit{Ab initio} potential energy points computed using LCCSD(T)/aug-cc-pV5Z and RCCSD(T)-F12/cc-pVTZ-F12 methods for three different geometries. Corresponding cuts through the constructed PES are given (solid curve).}
\label{fig:abinitio}
\end{figure}

We find the two methods to agree well for the geometries for which the convergence was achieved. In Fig. \ref{fig:abinitio}, we compare calculated \textit{ab initio} points for three representative geometries, given by $\beta=90^\circ$, $\beta=40^\circ$, and $\beta=10^\circ$. The convergence of the RCCSD(T)-F12/cc-pCVTZ-F12 calculation was not achieved for all values of $R$ for the angles $\beta=50^\mathrm{o}$ through $\beta=80^\mathrm{o}$, while for the collinear approach ($\beta=0^\circ$) the convergence of LCCSD(T)/aug-cc-pV5Z level of theory was generally achieved if $C_{\infty v}$ symmetry was explicitly given, unlike for $\beta=90^\mathrm{o}$ where either $C_{2v}$ or $C_{S}$ symmetry converged. 

We constructed the potential energy surface by varying the distance $R$ and angle $\beta$ between the approaching C atom and the center-of-mass of the CO$_2$ molecule while keeping the CO$_2$ molecule linear and in equilibrium. A non-equidistant grid in $R$, with the highest \textit{ab initio} point density around the global surface minimum, and an equidistant grid in angle $\beta=0^\circ \dots 90^\circ$ in steps of 10$^\circ$, were used. A total of 278 points were calculated per method. When constructing the PESs, we used the RCCSD(T)-F12/cc-pCVTZ-F12 points, except for the angles $\beta=50^\circ \dots 80^\circ$, for which we used the LCCSD(T)/aug-cc-pV5Z points in order to assure the convergence for all geometries.

\begin{figure}
\noindent\includegraphics[width=\columnwidth]{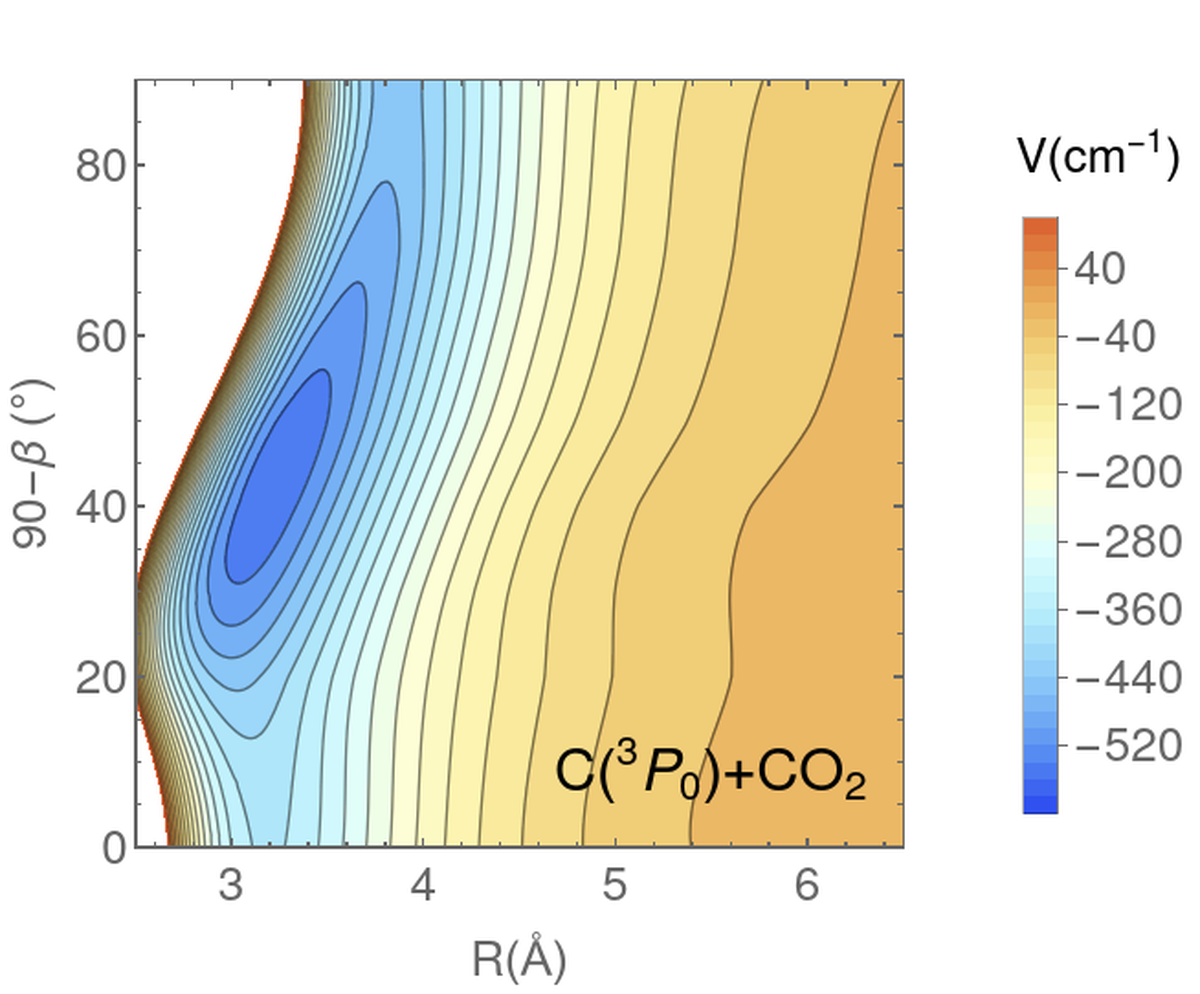}
\caption{A contour plot of the constructed C($^3P$)-CO$_2$ potential energy surface. The collinear approach ($\beta=0^\circ$) and the T-approach ($\beta=90^\circ$) correspond to the top and bottom of the $y$-axis, respectively.}
\label{fig:pes-contour}
\end{figure}

The final potential energy surface was constructed by fitting the \textit{ab initio} points to a functional form introduced by \citet{1999JChPh.110.3785B} and further adapted by \citet{2012CPL...549...12S} (Fig. \ref{fig:pes-contour}). We note that the counterpoise technique by \citep{boys1970calculation}, commonly used to compensate for the basis set superposition errors, did not result in significant modification of the final surface. 
All calculated interaction energies and fitting parameters are available on request.

\subsection{Collisional Dynamics}

We computed state-to-state quantum mechanical elastic and inelastic cross sections for two isotopes of impacting carbon atoms: $^{12}$C($^3P$) + CO$_2(j) \rightarrow$ $^{12}$C($^3P$) + CO$_2(j')$, and $^{13}$C($^3P$) + CO$_2(j) \rightarrow$ $^{13}$C($^3P$) + CO$_2(j')$, where the initial and final rotational quantum numbers of the CO$_2$ molecule are labeled as $j$ and $j'$, respectively. Throughout the text, the isotope $^{12}$C will be assumed if the atomic mass superscript is omitted in the notation.
The scattering process was modeled using time-independent coupled-channel formalism with the rigid rotor approximation applied to CO$_2(j)$ molecule according to \citet{arthurs1960theory} theory.
The calculation was carried out using MOLSCAT code \citep{MOLSCAT} with the atom-rigid rotor interaction described using the analytic representation of our newly constructed PES.

For each of the two isotopes of the impacting C atom, namely $^{12}$C and $^{13}$C, the scattering cross sections were calculated for a total of 41 energy points between $E_\mathrm{col}=240$ cm$^{-1}$ [$2.976\times10^{-2}$ eV] and $E_\mathrm{col}=40,240$ cm$^{-1}$ [4.9891 eV] in steps of 10$^3$ cm$^{-1}$ [0.124 eV].
We confirmed that the numerical convergence was achieved at all energy points by performing convergence tests with respect to all relevant parameters, including the number of partial waves, basis size given by the largest value of the rotational quantum number $j_\mathrm{max}$, and numerical integration parameters\citep{MOLSCAT}.


As in case of O($^3P$)+CO$_2$ scattering \citep{2020MNRAS.491.5650G}, the elastic cross sections were found to converge to within 5\% for a basis set cutoff in the rotational quantum number set as low as $j_\mathrm{max}=30$.
A significantly larger cutoff value was required to ensure the convergence of inelastic cross sections. To keep the computation time and memory requirements realistic, we used the centrifugal sudden (CS8) approximation  \citep{mcguire1974quantum}, where the dipole transitions between the rotational states whose difference is larger than $\Delta j=8$ were not computed.
A satisfactory (relative uncertainties less than 5\%) convergence of the inelastic cross sections was achieved for $j_\mathrm{max}=90$ and above. The convergence with respect to the total rotational quantum number of the complex, $J_\mathrm{max}$, was achieved for $J_\mathrm{max} = 200-1300$, with more partial waves needed at higher collision energies. In all cases, numerical integration of the coupled-channel system was carried out using modified log-derivative method \citep{1986JChPh..85.6425M} replaced with the hybrid modified log-derivative/Airy (LDA) propagator \citep{alexander1987stable} at large separation projectile-target separations, where the Airy propagator was set up to automatically detect integration region boundaries. In such approach, using the faster LDA propagator did not result in an observable loss of numerical accuracy. The parameters used to obtain the reported results are given in Table \ref{table_molscat}.

\section{Results}

\subsection{Elastic and inelastic cross sections}

\begin{figure}
  \centering
  \noindent\includegraphics[width=\columnwidth]{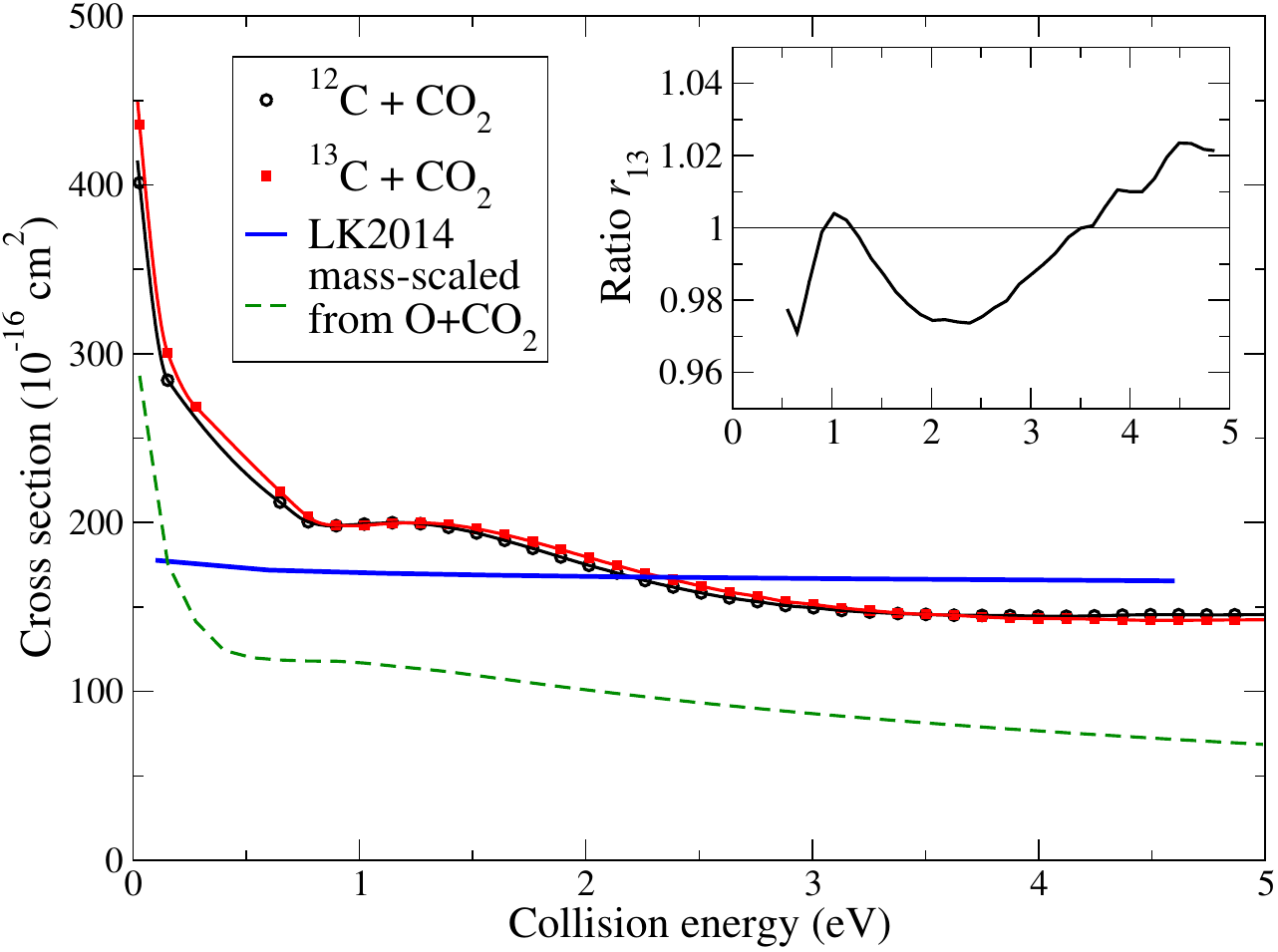}
  \caption{Elastic cross sections $\sigma_{j=0}^\mathrm{el}$ for $^{12}$C($^3P$) + CO$_2$($j$=0) and $^{13}$C($^3P$) + CO$_2$($j$=0) scattering as a function of the collision energy $E$. The elastic cross section of \citet{2014ApJ...790...98L} (LK2014) and mass-scaled from O($^3P$)+CO$_2$ \citep{2020MNRAS.491.5650G} are given for $^{12}$C($^3P$) + CO$_2$($j$=0). \textit{Inset:} The ratio of the elastic cross sections (this work) for the two isotopes, $r_{13} = \sigma_{j=0}^\mathrm{el}[^{12}\mathrm{C}] / \sigma_{j=0}^\mathrm{el}[^{13}\mathrm{C}]$. }
  \label{fig:elastic-cs-isotopes}
\end{figure}

\begin{table}
\caption{Numerical parameters used in MOLSCAT calculation.}
\centering
\begin{tabular}{l r}
\hline
 Parameter          & Value  \\
\hline
  $j_\mathrm{max}$  & 90            \\
  $J_\mathrm{max}$  & 200-1300      \\
  JZCSMX            & 8             \\
  RMIN (\AA{})      & 1.7$^a$       \\
  RMAX (\AA{})      & 15$^a$        \\
  RVFAC             & 1.3           \\
  STEPS             & 12            \\
\hline
\multicolumn{2}{l}{$^{a}$Automatic scaling algorithms used.}
\end{tabular}
\label{table_molscat}
\end{table}

\begin{figure}
  \centering
  \noindent\includegraphics[width=\columnwidth]{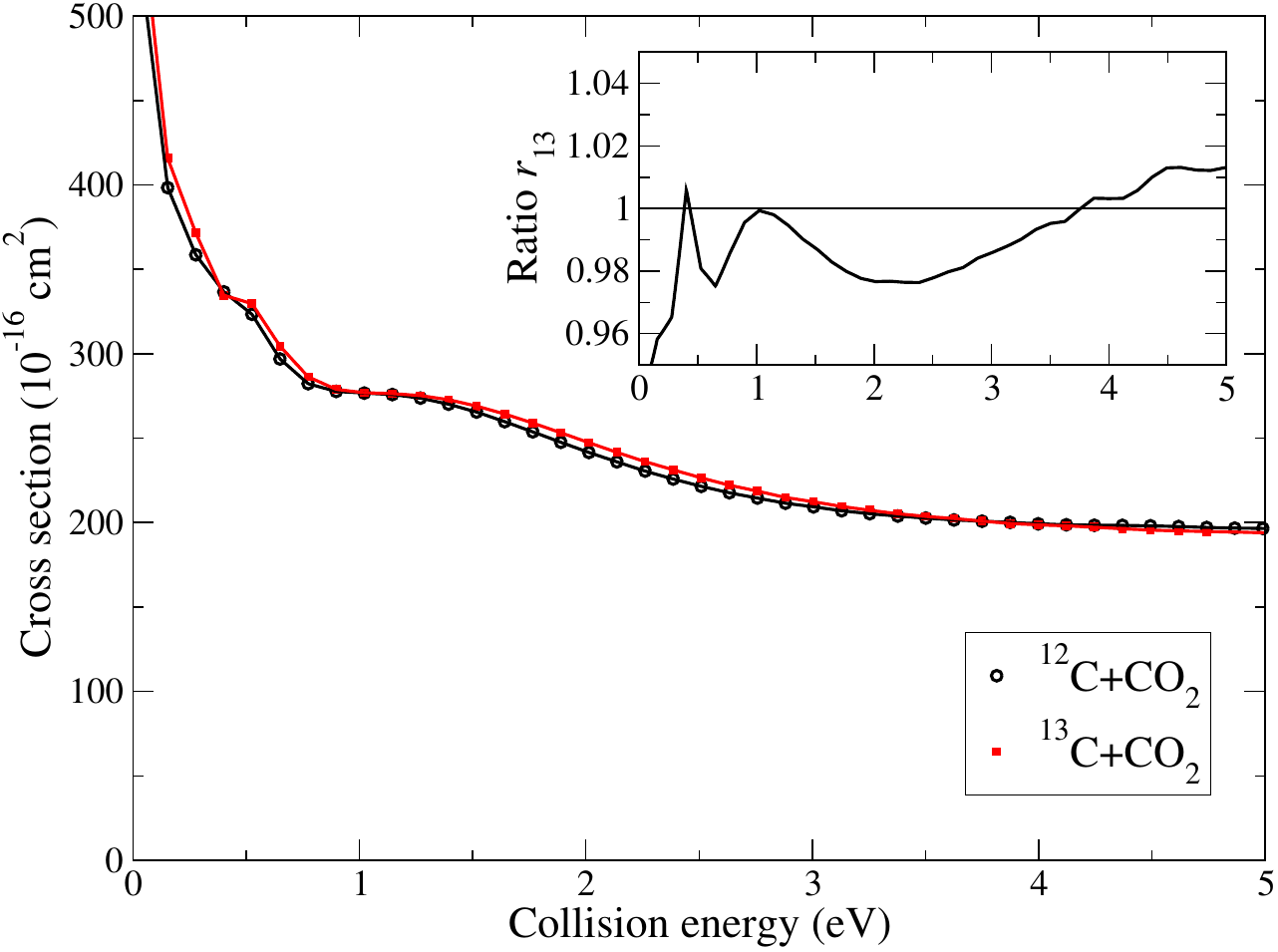}
  \caption{Total cross sections $\sigma_{j=0}^\mathrm{tot}$ for $^{12}$C($^3P$) + CO$_2$($j$=0) and $^{13}$C($^3P$) + CO$_2$($j$=0) scattering as a function of the collision energy $E$. \textit{Inset:} The ratio of the total cross sections for the two isotopes, $r_{13} = \sigma_{j=0}^\mathrm{tot}[^{12}\mathrm{C}] / \sigma_{j=0}^\mathrm{tot}[^{13}\mathrm{C}]$.}
  \label{fig:total-cs-isotopes}
\end{figure}

\begin{table}
\caption{Integral cross sections for $^{12,13}$C($^3P$) + CO$_2$($j$=0) scattering (units of $10^{-16}$ cm$^2$) for selected collision energies.}
\centering
\begin{tabular}{|r|rr|rr|}
\hline
         & \multicolumn{2}{c}{$^{12}$C($^3P$) + CO$_2$} & \multicolumn{2}{c}{$^{13}$C($^3P$) + CO$_2$} \\
$E$ (eV)	& elastic	& total	                           & elastic	& total \\
\hline
0.03	   & 401.33	& 535.78 & 435.62 & 569.01 \\
0.15	   & 284.40	& 398.40 & 300.37	 & 415.75 \\
0.28	   & 257.28	& 358.62 & 268.59	 & 371.50 \\
0.40	   & 242.28	& 336.67 & 239.05	 & 334.65 \\
0.53	   & 234.76	& 323.37 & 239.86	 & 329.70 \\
0.65	   & 212.62	& 296.95 & 219.20	 & 304.50 \\
0.77	   & 201.44	& 282.30 & 203.77	 & 286.37 \\
0.90	   & 198.72	& 277.67 & 198.06 & 278.92 \\
1.02	   & 199.60	& 276.73 & 200.13 & 276.92 \\
1.27	   & 199.91	& 273.74 & 200.45 & 275.23 \\
1.52	   & 194.63	& 265.50 & 197.10 & 269.02 \\
1.77	   & 180.66	& 253.76 & 189.51 & 258.98 \\
2.01	   & 175.50	& 241.56 & 180.14 & 247.34 \\
2.26	   & 166.30	& 230.54 & 171.04 & 236.11 \\
2.51	   & 159.16	& 221.50 & 163.03 & 226.50 \\
2.76	   & 153.76	& 214.55	 & 157.04 & 	218.68 \\
3.01	   & 150.20	& 209.35	 & 152.13 & 	212.34 \\
3.25	   & 147.56	& 205.28	 & 148.48 & 	207.32 \\
3.50	   & 146.14	& 202.68	 & 146.12 & 	203.66 \\
3.75	   & 145.41	& 200.85	 & 144.50 & 	200.91 \\
4.00	   & 146.85	& 199.30	 & 143.65 & 	198.68 \\
4.25	   & 146.01	& 198.42	 & 142.32 & 	197.28 \\
4.49	   & 146.25	& 198.11	 & 144.34 & 	195.57 \\
4.74	   & 147.39	& 197.14	 & 142.73 & 	194.75 \\
4.87	   & 146.31	& 196.75	 & 142.73 & 	194.36 \\
\hline
\hline
\end{tabular}
\label{table_cs}
\end{table}

For the title processes, following the approach described in the previous section, we calculated the state-to-state cross sections $\sigma_{j,j'}$ as a function of the collision energy $E$ given in the center-of-mass frame for all combinations of initial and final rotational quantum numbers $j$ and $j'$.
For each $j$ and the collision energy $E$, the elastic scattering cross section is simply $\sigma_{j}^\mathrm{el}(E) = \sigma_{j,j}(E)$ and the total scattering cross section is given by the sum over all rotational states of the CO$_2$ molecule after the collision
\begin{equation}
  \sigma_{j}^\mathrm{tot}(E) = \sum_{j'} \sigma_{j,j'}(E) ,
\end{equation}
where $j'=0 \ldots j_\mathrm{max}$.

In Figs. \ref{fig:elastic-cs-isotopes} and \ref{fig:total-cs-isotopes} the elastic and total scattering cross sections for the initially non-rotating CO$_2$ molecules, $\sigma_{j=0}^\mathrm{el}$ and $\sigma_{j}^\mathrm{tot}$, respectively, are shown as a function of the collision energy $E=0.03$ eV to 5 eV for the two isotopes of impacting carbon atoms as well as tabulated for selected energies in Table \ref{table_cs}.
The computed elastic cross sections show significant differences from the velocity-dependent cross sections available in the literature. When compared to the elastic cross sections of \citet{2014ApJ...790...98L} (LK2014), the reported cross sections exhibit stronger dependence on the collision energy. The difference is more pronounced for energies below 1.5 eV, where our cross sections are about 20\% (for $E$=1.5 eV) to more than 100\% (for $E$=0.07 eV) larger than LK2014 values. At higher energies, at $E>3$ eV, our cross sections converge to a constant value that is about 10\% smaller than LK2014 cross sections.
Moreover, within the considered collision energy range, the reported elastic cross sections are about 80\% greater than predicted by simple mass-scaling from the O($^3P$)+CO$_2$ \citep{2021Icar..36014371L,2020MNRAS.491.5650G}.

To simplify the comparison between the two isotopes of carbon atoms, the energy-dependent ratios of their elastic and total cross sections, $r_{13}^{\mathrm{el,tot}} = \sigma_{j=0}^\mathrm{el,tot}[^{12}\mathrm{C}] / \sigma_{j=0}^\mathrm{el,tot}[^{13}\mathrm{C}]$, are given in Figs. \ref{fig:elastic-cs-isotopes} and \ref{fig:total-cs-isotopes} (inset), respectively. They exhibit a very similar dependence on the collision energy, indicating that the elastic scattering plays a dominant role. Within three energy ranges, namely 0.03 eV < $E$ < 0.35 eV, 0.5 eV < $E$ < 0.8 eV, and 1.5 eV < $E$ < 3 eV, cross sections for the impacting $^{12}$C are 2\% or more smaller than for $^{13}$C isotope. In contrast, for $E>3.6$ eV, their ratio changes in favor of $^{12}$C isotope.

\begin{figure}
  \centering
  \noindent\includegraphics[width=\columnwidth]{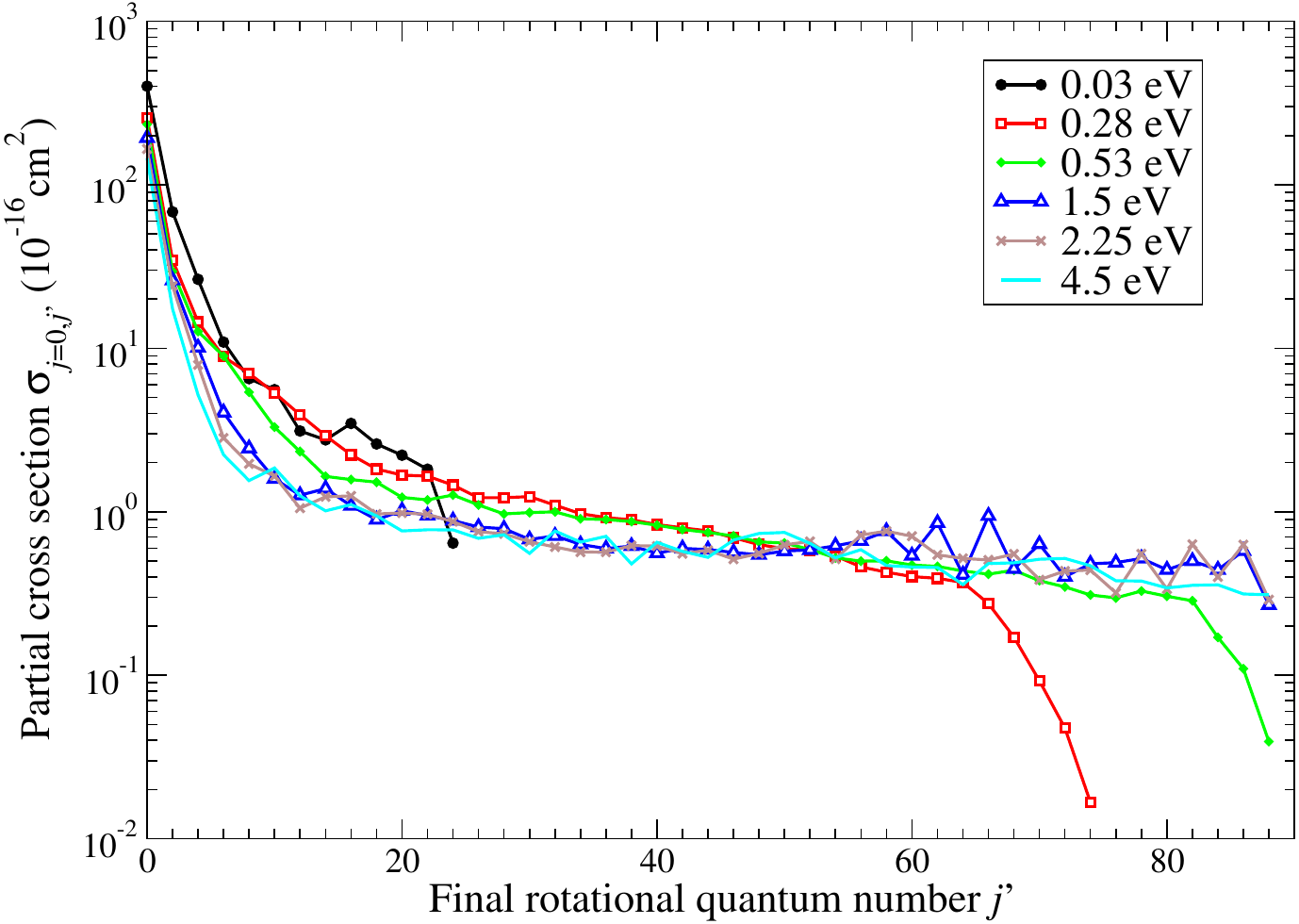}
  \caption{Partial cross sections $\sigma_{j=0,j'}$ for $^{12}$C($^3P$) + CO$_2(j=0) \rightarrow$ $^{12}$C($^3P$) + CO$_2(j')$ scattering given for selected collision energies as a function of rotational quantum number $j'$.}
  \label{fig:inelastic-cs}
\end{figure}

Inelastic cross sections $\sigma_{j=0,j'}(E)$ for $^{12}$C($^3P$) + CO$_2(j=0) \rightarrow$ $^{12}$C($^3P$) + CO$_2(j')$ scattering are shown in dependence of rotational quantum number $j'$ in Fig. \ref{fig:inelastic-cs} for selected collision energies $E$. This scenario corresponds to translationally hot carbon atoms colliding with thermal CO$_2$ gas. Here, at low collision energies, $E<0.3$ eV, the inelastic cross sections are larger by up to an order of magnitude for low values of $j'$, indicating that rotational excitations are more efficient than at higher collision energies, in agreement with general predictions from the scattering theory.

\begin{table*}
\caption{State-to-state cross sections for $^{12}$C($^3P$) + CO$_2$($j$=0) scattering (units of $10^{-16}$ cm$^2$) for selected collision energies.}
\centering
\begin{tabular}{r|rrrr rrrrr}
\hline
        & \multicolumn{9}{c}{Center-of-mass collision energy $E$ (eV)} \\
 $j'$   & 0.03   & 0.28  & 0.53  & 0.77  & 1.02 & 1.5  & 2.26 & 3.5 & 4.5 \\
\hline
0 & 401.324 & 256.775 & 234.487 & 200.453 & 199.154 & 193.849 & 165.656 & 145.637 & 145.467 \\
2&68.3663&34.6633&29.5356&27.1094&26.6361&26.047&24.0761&20.0731&17.3975 \\
4&26.4247&14.491&12.6783&12.7031&12.0122&10.1052&7.91625&6.00399&5.18549 \\
6&10.9098&8.88699&8.95512&7.6208&6.03346&4.05253&2.82757&2.23402&2.23083 \\
8&6.52679&7.01532&5.38974&3.78105&3.04357&2.44427&1.96517&1.60265&1.55226 \\
10&5.58287&5.3432&3.29722&2.21903&1.79568&1.59195&1.65628&1.97337&1.85149 \\
12&3.12262&3.91948&2.33609&1.92948&1.66758&1.26329&1.05273&1.24388&1.25211 \\
14&2.75617&2.9195&1.64655&1.4618&1.45524&1.38627&1.23449&1.11584&1.01313 \\
16&3.47292&2.2306&1.57411&1.24859&1.08966&1.08873&1.25008&1.17158&1.10766 \\
18&2.60125&1.82583&1.51856&1.33933&1.12798&0.894677&0.970187&0.941696&0.944952 \\
20&2.21932&1.67722&1.22527&1.18633&1.13796&1.01698&0.982231&0.765726&0.763308 \\
22&1.82226&1.65229&1.18317&0.95754&0.925198&0.947543&0.965936&0.836626&0.776433  \\
24&0.64374&1.45597&1.26635&0.99916&0.869824&0.885191&0.871104&0.710202&0.775156 \\
26&&1.22009&1.10095&1.0165&0.94086&0.806275&0.756057&0.714643&0.688039  \\
28&&1.21936&0.970015&0.880804&0.858167&0.78822&0.733544&0.631119&0.723374  \\
30&&1.23765&0.989488&0.844094&0.772114&0.677686&0.65636&0.679467&0.556519 \\
32&&1.09768&0.99898&0.875506&0.775674&0.712077&0.609136&0.646915&0.763255 \\
34&&0.973927&0.905967&0.820596&0.741321&0.631615&0.566797&0.61629&0.652802 \\
36&&0.921907&0.897234&0.770033&0.712032&0.59541&0.56636&0.768978&0.707538 \\
38&&0.893776&0.870997&0.750954&0.657156&0.616715&0.621074&0.639651&0.480125 \\
40&&0.835861&0.827643&0.732733&0.642326&0.558326&0.617224&0.564043&0.650754 \\
42&&0.796883&0.779928&0.701964&0.613693&0.597033&0.554656&0.495692&0.572498 \\
44&&0.764573&0.747445&0.665384&0.634108&0.587729&0.582373&0.554466&0.526183 \\
46&&0.693603&0.712067&0.681205&0.590588&0.564234&0.51305&0.576965&0.671481 \\
48&&0.630077&0.654706&0.62518&0.592153&0.546247&0.555946&0.655523&0.737486 \\
50&&0.594434&0.644639&0.634405&0.576241&0.575339&0.627566&0.736257&0.747189 \\
52&&0.582099&0.588582&0.592116&0.648283&0.5846&0.658818&0.66313&0.640635 \\
54&&0.536851&0.522942&0.587642&0.538116&0.619776&0.513755&0.669838&0.523245 \\
56&&0.460184&0.499495&0.547018&0.684895&0.663933&0.72041&0.736676&0.586429 \\
58&&0.427049&0.501287&0.537004&0.507546&0.760488&0.760985&0.578092&0.469333 \\
60&&0.40175&0.472843&0.57758&0.757063&0.53975&0.709113&0.496145&0.459552 \\
62&&0.391413&0.463544&0.503444&0.440138&0.854023&0.545465&0.413908&0.456024 \\
64&&0.369847&0.434698&0.518835&0.798318&0.412657&0.518797&0.369499&0.358804 \\
66&&0.274164&0.415146&0.372572&0.396237&0.943628&0.507418&0.420325&0.48291 \\
68&&0.171155&0.439141&0.545257&0.704827&0.447355&0.55177&0.558242&0.487159 \\
70&&0.0925556&0.37876&0.38209&0.36033&0.634371&0.383843&0.374619&0.512453 \\
72&&0.0476278&0.346935&0.494899&0.589563&0.400621&0.434618&0.323787&0.518411 \\
74&&0.0167521&0.310003&0.340696&0.328313&0.480581&0.440139&0.514467&0.468069 \\
76&&&0.297422&0.479285&0.56593&0.488623&0.319565&0.424642&0.378858 \\
78&&&0.327916&0.30213&0.39251&0.518561&0.553353&0.585903&0.377246 \\
80&&&0.304192&0.524157&0.566743&0.442019&0.336872&0.453667&0.342414 \\
82&&&0.285694&0.41838&0.420498&0.499584&0.632724&0.299095&0.354913 \\
84&&&0.170565&0.326018&0.465344&0.442403&0.399575&0.375898&0.357701 \\
86&&&0.109612&0.484479&0.545315&0.576273&0.62991&0.366574&0.314384 \\
88&&&0.0393414&0.165967&0.205915&0.269364&0.29006&0.332944&0.310817 \\

\hline
\end{tabular}
\label{table_cs1}
\end{table*}

At low collision energies, $E<0.5$ eV, the basis set used is sufficiently large to fully capture all rotational excitations (the highest excited rotational quantum number $j'$ is smaller than $j_\mathrm{max}=90$). At higher collision energies, the highest rotational excitations are not captured but the numerical solver converged in all cases due to the fact that the contribution from $j'>j_\mathrm{max}$ is not significant. Based on the procedure described by \citet{2020MNRAS.491.5650G}, the uncertainties in the cross sections increase with the collision energy, from about 1-2\% for $20<j'<40$ at $0.5<E<1$ eV to about 10\% for $j'>80$ at $E>4$ eV.
A complete list of state-to-state cross sections for selected collision energies of interest for planetary aeronomy is given in Table \ref{table_cs1}. The complete datasets are available on Zenodo archive \citep{mgacesa_2023_8224848}.

\subsection{Differential cross sections}

\begin{figure}
  \centering
  \noindent\includegraphics[width=\columnwidth]{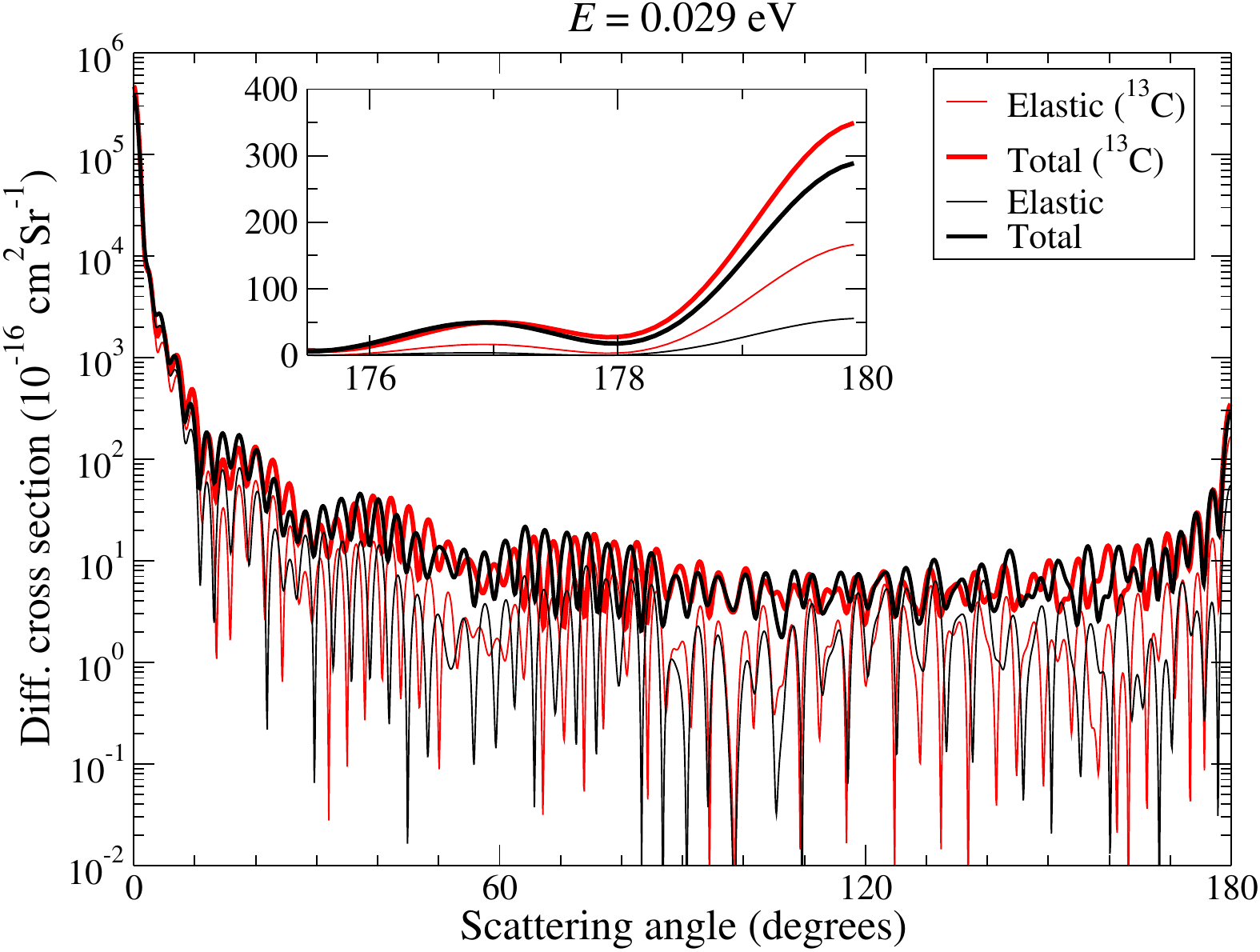}
  \caption{Differential cross sections for a low collision energy ($E$=0.0297 eV) in dependence of the scattering angle. Elastic, $Q_{j=0,j'=0}(\theta,E)$ (thin curves), and total, $Q^{\mathrm{tot}}_{j=0}(\theta,E)$ (thick curves), DCSs are shown for $^{12,13}$C+CO$_2(j=0)$ scattering. \textit{Inset:} Zoom in on the large scattering angles characteristic of backscattering.}
  \label{fig:DCS_iE1}
\end{figure}
\begin{figure}
  \centering
  \noindent\includegraphics[width=\columnwidth]{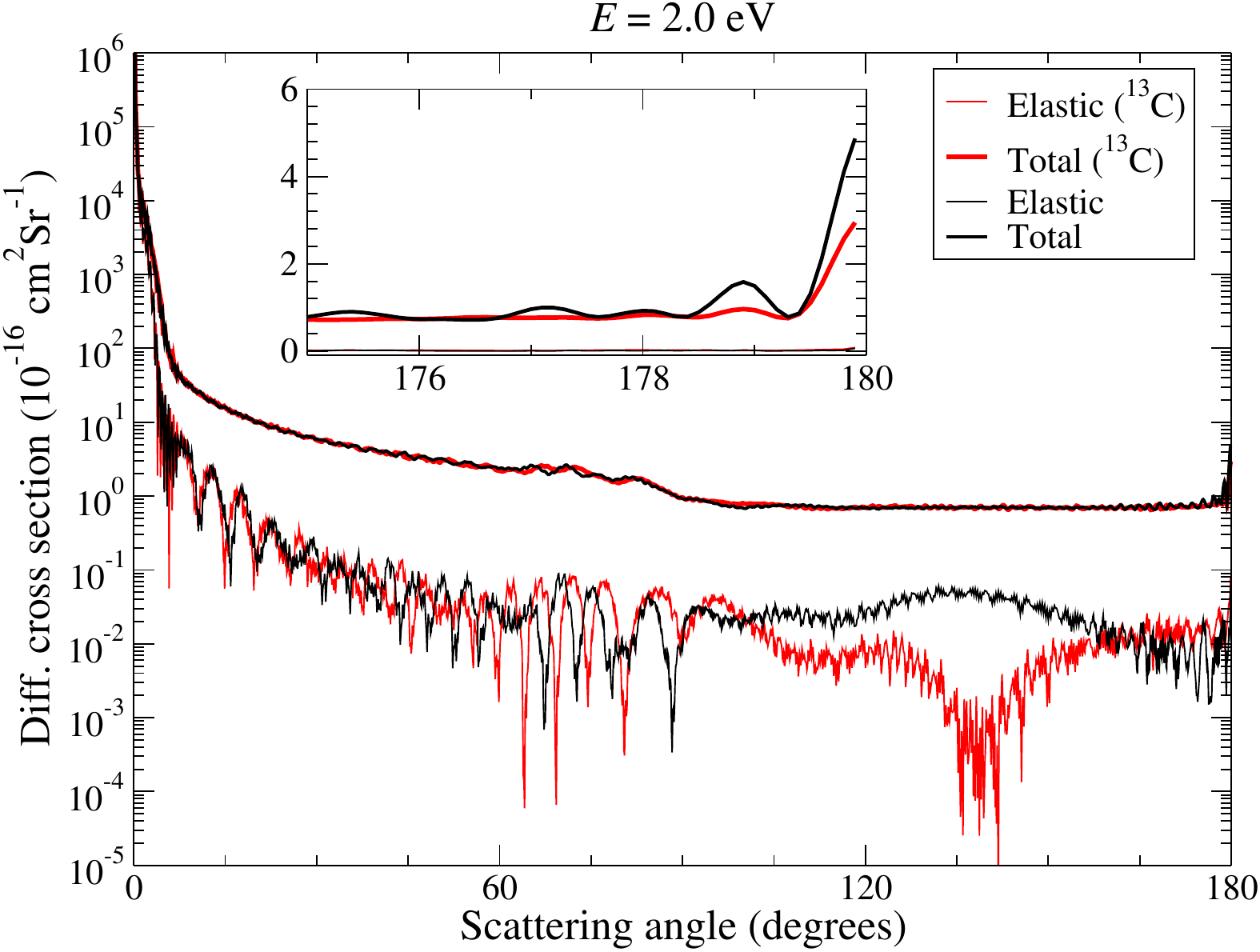}
  \caption{As above, for the collision energy $E$=2 eV.}
  \label{fig:DCS_iE17}
\end{figure}

We constructed state-to-state differential cross sections (DCSs), $Q_{j,j'}(\theta,E) = {\partial \sigma_{j,j'}(\theta,E)}/{\partial \Omega}$, for all considered collision energies using numerically evaluated \textit{S}-matrices. Here, $d\Omega = \sin \theta d \theta d \phi$ is the solid angle element and $\theta$ is the scattering angle, defined as the relative angle between an incoming and outgoing trajectory of the impacting C($^3P$) atom. In this notation, $\theta=0^\circ$ corresponds to forward scattering without deflection, $\theta=90^\circ$ to sideways scattering, and $\theta=180^\circ$ to complete backscattering.
The matrices were processed using the utility code \texttt{dcs\_save.f} included with the MOLSCAT v14 \citep{MOLSCAT} release.
For each initial state $j$, we additionally constructed the total differential cross section as a sum over all open channels populated in a collision, $Q^{\mathrm{tot}}_{j}(\theta,E) = \sum_{j'} Q_{j,j'}(\theta,E)$.

Sample DCSs for two collision energies, 0.0297 eV and 2.5 eV, representative of low- and high-energy scattering, are illustrated in Figs. \ref{fig:DCS_iE1} and \ref{fig:DCS_iE17} for both carbon isotopes considered, respectively, and CO$_2$ in the initial state $j=0$.
The DCSs are strongly anisotropic, with the forward-scattering peak ($\theta<2^{\circ}$) contributing more than 90\% of the integral cross section, similar to other cases of atom-atom and atom-molecule scattering at eV energies \citep{1998JGR...10323393B,2000JGR...10524899K,2012GeoRL..3910203G}.
For backscattering ($\theta=180^{\circ}$), the cross sections are enhanced by a factor of 3 to 6, as compared to the large angle scattering, with almost the entire contribution coming from inelastic excitations (Figs. \ref{fig:DCS_iE1} and \ref{fig:DCS_iE17}, \textit{inset}). Moreover, the total DCSs for backscattering of $^{12}$C+CO$_2(j=0)$ are larger by nearly a factor of two than for $^{13}$C+CO$_2(j=0)$. A similar enhancement is present for other initial rotational states $j$ of CO$_2$, suggesting that different transport properties for the two isotopes may be present for geometries where the backscattering is important.
The total DCSs are nearly identical to the elastic ones for small scattering angles ($\theta<1^{\circ}$). The elastic DCSs rapidly decrease with the scattering angle and for $\theta>5^\circ$ become on average about 2 orders of magnitude smaller than the total DCSs for scattering energies larger than about 0.5 eV (Fig. \ref{fig:DCS_iE17}). The difference becomes smaller at low energies (Fig. \ref{fig:DCS_iE1}). Except for the forward- and backscattering, the differences in DCSs between the isotopes are more subtle and change with the scattering energy. A common trend present for higher scattering energies is that for $\theta>90^{\circ}$ the elastic DCS for the $^{13}$C scattering decreases with respect to the $^{12}$C scattering (see Fig. \ref{fig:DCS_iE17}), suggesting that the heavier isotope is preferentially scattered at smaller angles.
The complete datasets containing state-to-state DCSs are available online \citep{mgacesa_2023_8224848}.

\subsection{Momentum transfer cross sections}

Momentum transfer cross sections (MTCS; also called momentum transport or diffusion cross sections) are effective quantities suitable for transport studies, such as energy exchange, diffusion, and transport of a particle in a background gas at non-thermal conditions encountered in astrophysical environments and planetary atmospheres \citep{DALGARNO1962643,1964mtgl.book.....H,1997JASTP..59..107K}.

For the title scattering process, C+CO$_2(j) \rightarrow$ C+CO$_2(j')$, the state-to-state momentum transfer cross section (MTCS) for inelastic collisions is given by \citep{1978JChPh..68.1585P}
\begin{equation}
  \sigma^{\mathrm{mt}}_{j, j'}(E) = 2 \pi \int_0^{\pi} \mathrm{d} \theta \sin \theta
       \left(1 - \sqrt{ 1 - \gamma_{j'}} \cos \theta \right) Q_{j,j'}(\theta,E) \; ,
  \label{mtcs}
\end{equation}
where $\gamma_{j'} = \epsilon_{j'} / \epsilon$, is the ratio of $\epsilon_{j'}$ and $\epsilon$, the translational kinetic energies in the center-of-mass frame after and before the collision, respectively. 
We define the total MTCS for the initial state $j$ as
\begin{equation}
  \sigma^{\mathrm{mt,tot}}_{j}(E) = \sum_{j'} \sigma^{\mathrm{mt}}_{j, j'}(E) .
  \label{eq:mtcs_total}
\end{equation}
\begin{figure}
  \centering
  \noindent\includegraphics[width=0.95\columnwidth]{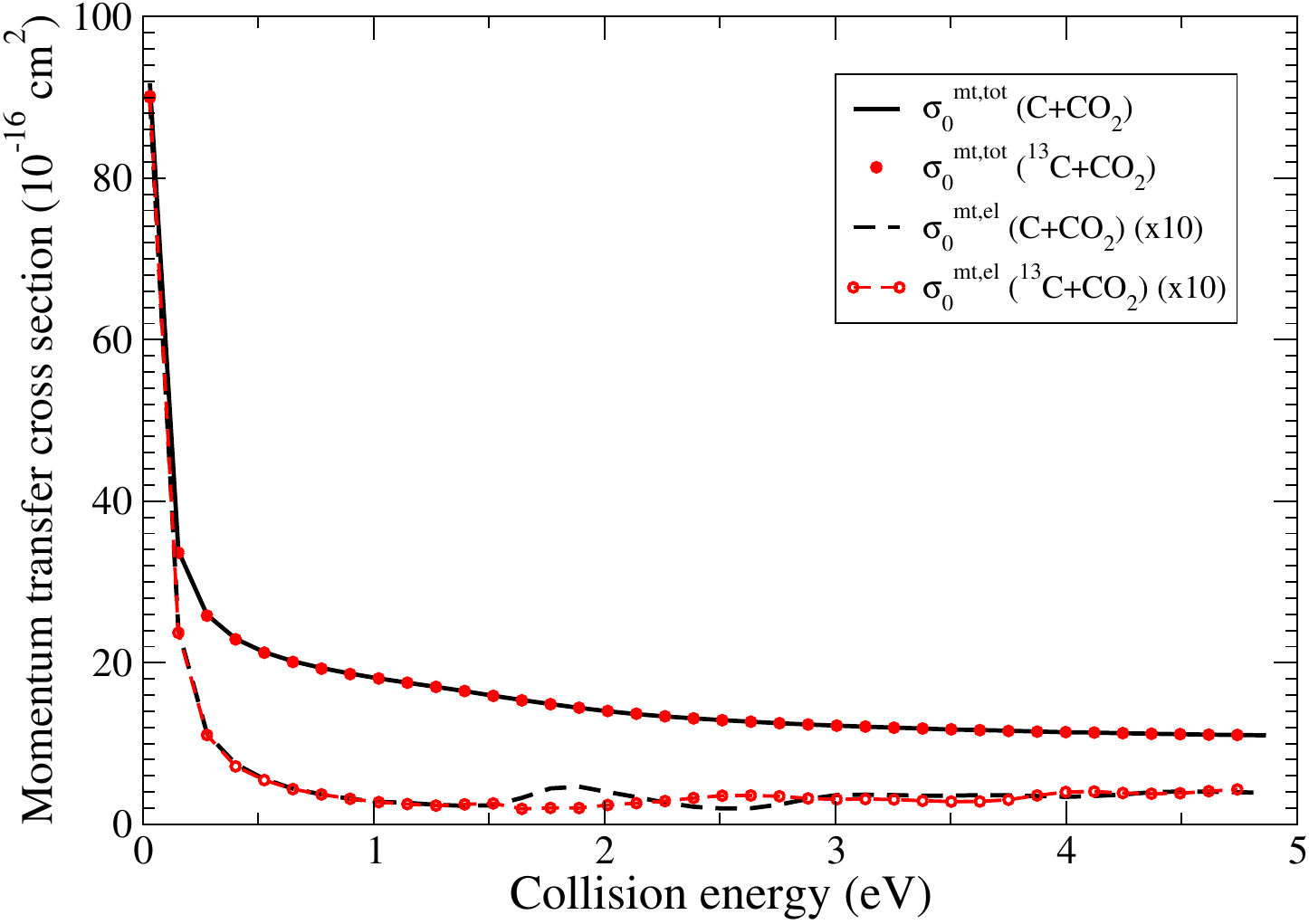}
  \caption{Total ($\sigma_{j=0}^{\mathrm{mt,tot}}$) and elastic ($\sigma^{\mathrm{mt}}_{j=0, j'=0}$) momentum transfer cross sections for $^{12,13}$C + CO$_2$($j=0$) as a function of the collision energy for the two considered isotopes of carbon. The elastic cross sections are multiplied by a factor of ten for visual comparison.}
  \label{fig:mtcs}
\end{figure}
Using the above expressions, we calculated the total and elastic MTCSs, $\sigma^{\mathrm{mt,tot}}_{j}$ and $\sigma^{\mathrm{mt,el}}_{j=0} = \sigma^{\mathrm{mt}}_{j=0, j'=0}$, respectively, for the range of collision energies considered for both isotopes of carbon (Fig. \ref{fig:mtcs}).

As observed for other atom-molecule pairs \citep{1978JChPh..68.1585P,2015JChPh.143e4303D,2020MNRAS.491.5650G,chhabra2023quantum}, the MTCSs at low energies are dominated by the elastic cross sections and rapidly decrease monotonically as the energy increases and converge to a near-constant value for the energies greater than about 3 eV. The elastic MTCSs exhibit oscillatory periodic features between about 1.5 eV and 3 eV, reflecting the character of the oscillations present in the elastic and total cross sections (Figs. \ref{fig:elastic-cs-isotopes} and \ref{fig:total-cs-isotopes}). The oscillations appear to be out of phase for the two carbon isotopes, resulting in a relative difference of up to a factor of two around the peaks at about 1.8 eV and 2.6 eV. The oscillations are not present in total MTCSs due to the summation over the state-to-state contributions.

\section{Summary and Discussion}

In this paper, we have carried out non-reactive quantum-mechanical scattering calculations for $^{12}$C($^3P$) + CO$_2$ and $^{13}$C($^3P$) + CO$_2$ pairs at thermal and superthermal collision energies up to 5 eV in the center-of-mass frame, using the ground state electronic potential energy surface computed in this work. The PES was constructed in a 2D planar geometry, with the CO$_2$ molecule in the ground electronic state and assumed to be linear with both C-O bonds fixed at the equilibrium, corresponding to the rigid rotor representation. Within this approximation, we computed velocity-dependent state-to-state cross sections, initial state-dependent elastic and total cross sections, and corresponding differential and momentum transfer cross sections. 

We have found that the computed cross sections differ significantly from the estimates obtained by mass-scaling the cross sections for different atom-molecule systems \citep{2014ApJ...790...98L,2021Icar..36014371L}, in particular at low energies, below 1 eV. When compared to the values obtained by mass-scaling the cross sections for O($^3P$)+CO$_2$ \citep{2020MNRAS.491.5650G}, the computed cross sections are more than two times larger, suggesting that the differences in the electronic structure between these complexes play an important role in dynamics. 
Moreover, we have found that the state-specific elastic and total cross sections differ by up to 2\% for the two isotopes of carbon colliding with $^{12}$CO$_2$. The relative differences are the most significant below 0.7 eV and between 1.5 eV and 3 eV, where the $^{13}$C yields mostly larger cross sections, while the lighter isotopes yields a larger cross section at the collision energies above 4 eV. As observed for other atom-molecule pairs, the scattering is strongly anisotropic and the excitations of rotational degrees of freedom of the CO$_2$ molecule is significant. This is in agreement with \citet{2012JPCA..116...64Y} who concluded that in O$(^3P)$ + CO$_2$ collisions, about 40\% of the collision energy remained as the translation energy of the products, while a simple energy partitioning based on statistical distribution of the level energies gives an even larger estimate \citep{2018Icar..300..411F}. Such a high efficiency of the kinetic-to-internal energy transfer may be explained the formation of a metastable C$_2$O$_2$ complex during the collision \citep{2012JPCA..116...64Y}. 

The main contributions to the uncertainties in the computed cross sections come from the linear rigid rotor approximation of the CO$_2$ molecule and the PES used to model the C($^3P$)-CO$_2$ electronic interaction. The vibrational and rotational transitions have been studied extensively in O($^3P$)+CO$_2$ collisions at superthermal energies \citep{schatz1981quasiclassical,1982CPL....88..553M,2012JGRA..117.4310C,1992JChPh..96.2025U,2012ACP....12.9013F}. These and related studies agree that the vibrational excitations due to the collisions are significant, in particularly of the bending mode of CO$_2$, and likely the leading inelastic process at collision energies greater than about 4 eV \citep{schatz1981quasiclassical}. From the energy partitioning arguments, the CO$_2(010)$ bending mode has a low excitation energy of about 667 cm$^{-1}$ (about 960 K or 0.083 eV), followed by the symmetric stretching mode (1388 cm$^{-1}$ or 0.172 eV) and the asymmetric stretching mode (2349 cm$^{-1}$ or 0.291 eV), making its density of states the largest of the three. 
At lower energies, the vibrational transitions are likely to be at least an order of magnitude smaller than the rotational transitions \citep{1982CPL....88..553M,1992JChPh..96.2025U,2015JChPh.143e4303D}. A more extensive discussion is given in \citet{2020MNRAS.491.5650G}. 
Including the bending mode in the collisional dynamics would require at least a 3-D or, preferably, a full-dimensional PES(s) and their couplings, none of which are currently available.

Finally, our study did not include possible reaction channels of the title process. Here, energetically permitted reaction product channels are isotope exchange [$^{12}$C($^3P$)+$^{13}$CO$_2 \rightarrow$ $^{13}$C($^3P$)+$^{12}$CO$_2$], and carbon atom abstraction [C+CO$_2$ -> CO+CO]. Neither process is particularly likely to occur, even though carbon isotope exchange could be relevant for isotope cycling in the middle and upper atmosphere of Mars. \citet{2012JPCA..116...64Y} studied these reaction channels for a hot O$(^3P)$ impacting on CO$_2$ and concluded that the oxygen isotope exchange reaction yield is less than 1.6\%, while the non-reactive scattering yield is greater than 98.4\%. Similarly, our earlier computational study of O$(^3P)$+H$_2$ reactive scattering \citep{2014JChPh.141p4324G} over a similar collision energy range predicted that reactions occur in less than 4\% of all processes, in good agreement with earlier results.

The computed scattering cross sections are directly applicable to the transport of superthermal atomic carbon encountered in dense astrophysical gases, such as CO$_2$-rich atmospheres of Mars \citep{Wordsworth_2013,jakosky2023mars} and Venus \citep{gillmann2022long}, exoplanetary atmospheres \citep{2023Natur.614..649J}, insterstellar molecular clouds \citep{sandford2020prebiotic}, and photodissociation dominated regions \citep{wolfire2022photodissociation}.

\section*{Acknowledgements}

MG was partially funded by the NASA MAVEN mission grant and Khalifa University grant \#8474000362-KU-FSU-2021. The author is grateful to R. Lillis, J. Fox, R. Yelle, and H. Ali Lulu for helpful suggestions and discussion.

\section*{Data Availability}

The data underlying this article are available in Zenodo (doi:10.5281/zenodo.8224848), at \url{https://doi.org/10.5281/zenodo.8224848}. The data will be shared on reasonable request to the corresponding author.




\bibliographystyle{mnras}
\bibliography{refs_CCO2} 




%
%


\bsp	
\label{lastpage}
\end{document}